\documentstyle[sprocl,epsfig]{article}

\def\Journal#1#2#3#4{{#1} {\bf #2}, #3 (#4)}

\def\NPA{{\em Nucl. Phys.} A}
\def\PLB{{\em Phys. Lett.}  B}

\def\PRL{\em Phys. Rev. Lett.}
\def\PRC{{\em Phys. Rev.} C}
\def\PRD{{\em Phys. Rev.} D}

\begin{document}

\title{Detailed analysis of $\eta$ production in proton-proton collisions}

\author{A. \v{S}varc, S. Ceci}

\address{Rudjer Bo\v{s}kovic Institute, Bijeni\v{c}ka c. 54, 10000 Zagreb, Croatia,
\\E-mail: svarc@rudjer.irb.hr, ceci@rudjer.irb.hr}

\maketitle


Coupled-channel, multiresonance partial wave analysis (PWA),
developed and used by Carnegie-Mellon-Berkeley analyses group
(CMU-LBL 79) \cite{Cut79} has been recently very frequently used,
and agreed as a potentially safe tool for extracting N$^*$
resonance parameters \cite{Man92,Sva99,Ben00,Day00}. The newly
formed subsection for resonance parameter analyses of Baryon
Resonance Analysis Group (BRAG) \cite{BRAGweb} has decided to
re-evaluate relatively old analyses \cite{Cut79,Hoe83}, what has
been suggested even  by their own authors recently \cite{Hoe01}.
BRAG has chosen three independent analyses which repeat the
formalism suggested earlier, but with the use of new and improved
data \cite{BRAG}. They have come to a certain level of agreement
regarding the number of poles and their values. However, some of
the "Cutkosky" like analyses have been dropped out
\cite{Bat98,Krusche}, for technical reasons presumably.

However, one possible direction of analysis has been dropped
altogether. The question arises whether the obtained PWA can be
"relatively" safely taken as input to processes involving more then
two particles in the final state. Temporarily forgetting the expected
additional complications (initial and final state interactions,
off-mass shell behavior of two body amplitudes, etc.) this work tends
to estimate whether the present two body T-matrices can account for
the observables of a three body process $pp \rightarrow pp \eta$, very
carefully measured in Uppsala near the threshold \cite{Upp}. The
special attention has been given to understanding the apparently
inverted shape of the proton-eta differential cross section in the
final state \cite{Upp,Nef01}. The tendency of this work is not to
improve the two body fit ( which needs a lot of additional observables
even to be semi-reliable, ) but to see if the present T-matrices can
explain the 2 $\rightarrow$ 3 body processes without drastic
assumptions of the complications of the three body physics.

The first, and natural test of the reliability of the two body amplitudes appeared in the
carefully measured total and differential cross section for the process $pp \rightarrow pp \eta$
\cite{Upp}. We have developed a simple model based on the exchange of the lowest mesons
depicted in the following figure:

\begin{figure}[h]
\begin{center}
\epsfig{figure=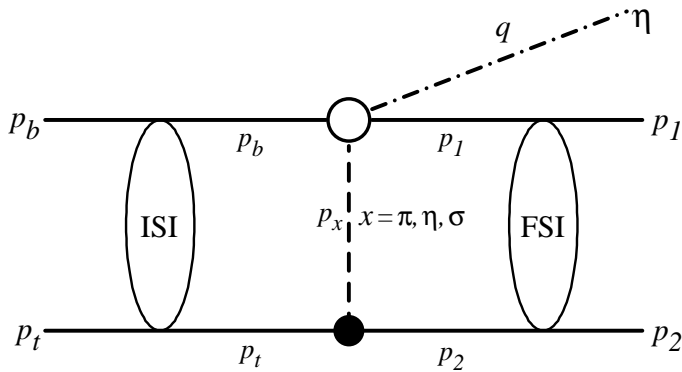,height=7cm,width=11cm}
\caption{The meson-exchange mechanism of $pp\rightarrow pp \eta$ reaction. The
initial and final state interactions are denoted as ISI and FSI
respectively.\label{fig:radish}}
\end{center}
\end{figure}

Unfortunately, there is a number of  models which claim to reproduce
the results, but differ among themselves drastically
\cite{Lag91,Vet91,Fal96b,Ged97,San98}, so it has been left to us to
show that our model \cite{Bat97}, reduced to the assumptions of the
mentioned models, gives a very similar result. The comparison is
successfully made and will be shown elsewhere. The main idea of this
presentation is to draw attention to the fact that the differential
$p\eta$ cross sections in a three body  process tends to show a
different curvature when compared to the two body $\pi^- p \rightarrow
\eta n$ process which should dominate the process \cite{Upp}. In spite
of the additional uncertainties  of the processes like ISI, FSI and
off-mass shell extrapolation of two body amplitudes, the effects
should be extremely high, and acting in the same direction in order to
turn the slope of the differential cross section. The disturbing data
are shown in Fig.2.

It is to be expected that the two $\rightarrow$ three body process is
dominated by the two body proton-meson $\rightarrow$ proton $\eta$
amplitude, in the vicinity of the threshold in  particular. However,
it turns out that even the shape of the differential cross sections of
the impulse approximation two body process \cite{Cec01} and the
measured 2$\rightarrow$3 body processes are different. Let us just
mention that only higher partial waves (like D$_{13}$) can account for
the opposite curvature. Therefore, we are left with only two
possibilities:
   either the ISI, FSI and off-mass shell effects of the higher order processes are
   responsible for the discrepancy,
   or the D$_{13}$ partial wave is not confidently extracted in 2$\rightarrow$2 body
   processes.

\begin{figure}[h]
\begin{center}
\epsfig{figure=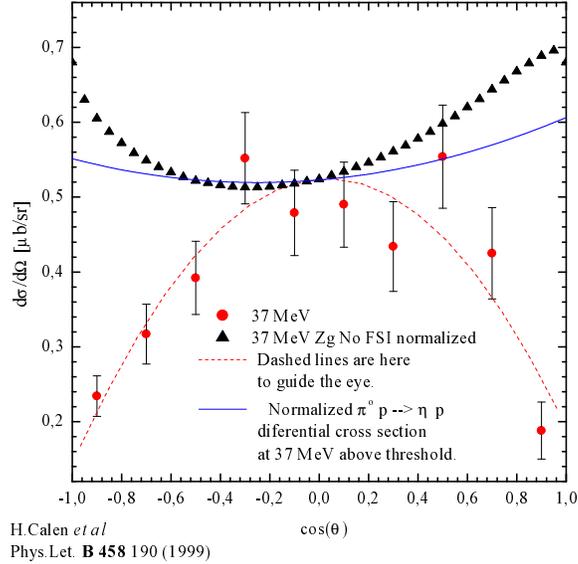,height=8cm,width=8cm}
\caption{The comparison of the experimental values (full dots with error bars) and the polynomial
 fit to them (dotted line) with $\pi^ - p\rightarrow \eta n$ at the comparable energy
(full triangles)  with the Zagreb calculation (full thin line).\label{fig:rad1}}
\end{center}
\end{figure}

     As it is obvious, the lowest partial wave in two body processes which can cause such a
       curvature are D$_{13}$, partial waves and they have been under close scrutiny at
       the Mainz workshop. Our feeling is that ISI, FSI and off-mass shell effects of higher
       order processes should be surprisingly strong to account for such a drastic change in
       the shape of the differential cross section. Therefore, there is an open possibility
       that something remains hidden in the D$_{13}$ two body partial wave which we have not
       been able to detect in two body processes. The future goal is to investigate all
       suggested possibilities, and see whether the $\rho$ exchanged meson domination in the
       hadron $\eta$ production  channel, which is quite an opened problem, can account for
       the apparent disagreement.  The results of further research will be reported.

\end{document}